\documentclass[11pt]{article}
\usepackage{fullpage}

\let\oldmarginpar\marginpar
\renewcommand\marginpar[1]{\-\oldmarginpar[\raggedleft\footnotesize #1]%
{\raggedright\footnotesize #1}}

\usepackage{caption}
\usepackage{subfig}
\usepackage{multirow, alltt}
\usepackage{rotating,algorithmic,amsmath}
\usepackage{amsmath,amsthm,amsfonts,amsgen,amstext,amssymb}
\usepackage{url}
\usepackage[ruled,vlined]{algorithm2e}
\usepackage{color,cite}
\usepackage{ifpdf}
\usepackage{fancyvrb}
\usepackage{listings}
\newcommand{\name}{\textsc{Bismarck}\xspace}
\newcommand{\ADB}{DBMS A\xspace}
\newcommand{\BDB}{DBMS B\xspace}
\newcommand{\chris}[1]{\textcolor{red}{#1}}

\newcommand{\eat}[1]{}

\newtheorem{example}{Example}[section]

\newcommand{\fullversion}[1]{#1}
\newcommand{\submissionversion}[1]{}

\def\compactify{\itemsep=0pt \topsep=0pt \partopsep=0pt \parsep=0pt}
\let\latexusecounter=\usecounter

\def\compactifytwo{\itemsep=-1pt \topsep=-1pt \partopsep=-2pt \parsep=-1pt \labelwidth=-2pt \leftmargin=-10pt} 
\newenvironment{CompactItemize}
 {\def\usecounter{\latexusecounter}
  \begin{itemize}
  \compactifytwo
  }
 {\end{itemize}\let\usecounter=\latexusecounter}

\begin{document}

\title{Towards a Unified Architecture for in-RDBMS Analytics}

\author{
Xixuan Feng\hspace{10mm}Arun Kumar\hspace{10mm}
Benjamin Recht\hspace{10mm}Christopher R\'{e}\vspace{0.2in}\\
Department of Computer Sciences\\
University of Wisconsin-Madison\\
\{xfeng, arun, brecht, chrisre\}@cs.wisc.edu
}
\date{}
\maketitle

\begin{abstract}
The increasing use of statistical data analysis in enterprise
applications has created an arms race among database vendors to offer
ever more sophisticated in-database analytics.  One challenge in this
race is that each new statistical technique must be implemented from
scratch in the RDBMS, which leads to a lengthy and complex development
process.  We argue that the root cause for this overhead is the lack
of a unified architecture for in-database analytics.  Our main
contribution in this work is to take a step towards such a unified
architecture.  A key benefit of our unified architecture is that
performance optimizations for analytics techniques can be studied
generically instead of an ad hoc, per-technique fashion.  In
particular, our technical contributions are theoretical and empirical
studies of two key factors that we found impact performance: the order
data is stored, and parallelization of computations on a single-node 
multicore RDBMS.  We demonstrate the feasibility of our architecture
by integrating several popular analytics techniques into two
commercial and one open-source RDBMS.  Our architecture requires
changes to only a few dozen lines of code to integrate a new
statistical technique.  We then compare our approach with the native
analytics tools offered by the commercial RDBMSes on various analytics
tasks, and validate that our approach achieves competitive or higher
performance, while still achieving the same quality.
\end{abstract}

\submissionversion{
\category{H.2}{Database Management}{Miscellaneous}
\keywords{Analytics, Convex Programming, Incremental Gradient Descent,
User-Defined Aggregate}
}

\section{Introduction}
There is an escalating arms race to bring sophisticated data analysis
techniques into enterprise applications. In the late 1990s and early
2000s, this brought a wave of data mining toolkits into the
RDBMS. Several major vendors are again making an effort toward
sophisticated in-database analytics with both open source efforts,
e.g., the MADlib platform~\cite{DBLP:journals/pvldb/CohenDDHW09}, and several projects
at major database vendors. In our conversations with engineers from
Oracle~\cite{orac:personal:communication} and
EMC Greenplum~\cite{gp:personal:communication}, we learned that a key
bottleneck in this arms race is that each new data analytics technique
requires several ad hoc steps: a new solver is employed that has new
memory requirements, new data access methods, etc. As a result, there
is little code reuse across different algorithms, slowing the
development effort. Thus, it would be a boon to the database industry
if one could devise a {\em single architecture} that was capable of
processing many data analytics techniques. An ideal architecture would
leverage as many of the existing code paths in the database as
possible as such code paths are likely to be maintained and optimized
as the RDBMS code evolves to new platforms.

To find this common architecture, we begin with an observation from
the mathematical programming community that has been exploited in
recent years by both the statistics and machine learning communities:
many common data analytics tasks can be framed as {\em convex
  programming problems}~\cite{Hast:Tibs:Frie:2001,boyd:cvx}. Examples
of such convex programming problems include support vector machines,
least squares and logistic regression, conditional random fields,
graphical models, control theoretic models, and many more. It is hard
to overstate the impact of this observation on data analysis theory:
rather than studying properties of each new model, researchers in this
area are able to unify their algorithmic and theoretical studies. In
particular, convex programming problems are attractive as local
solutions are always globally optimal, and one can find local
solutions via a standard suite of well-established and analyzed
algorithms. Thus, convex programming is a natural starting point for a
unified analytics architecture.

\begin{figure*}
\centering
\submissionversion{
\begin{tabular}{cc}
 \parbox[c]{2.7in}{
 \includegraphics[width=2.8in]{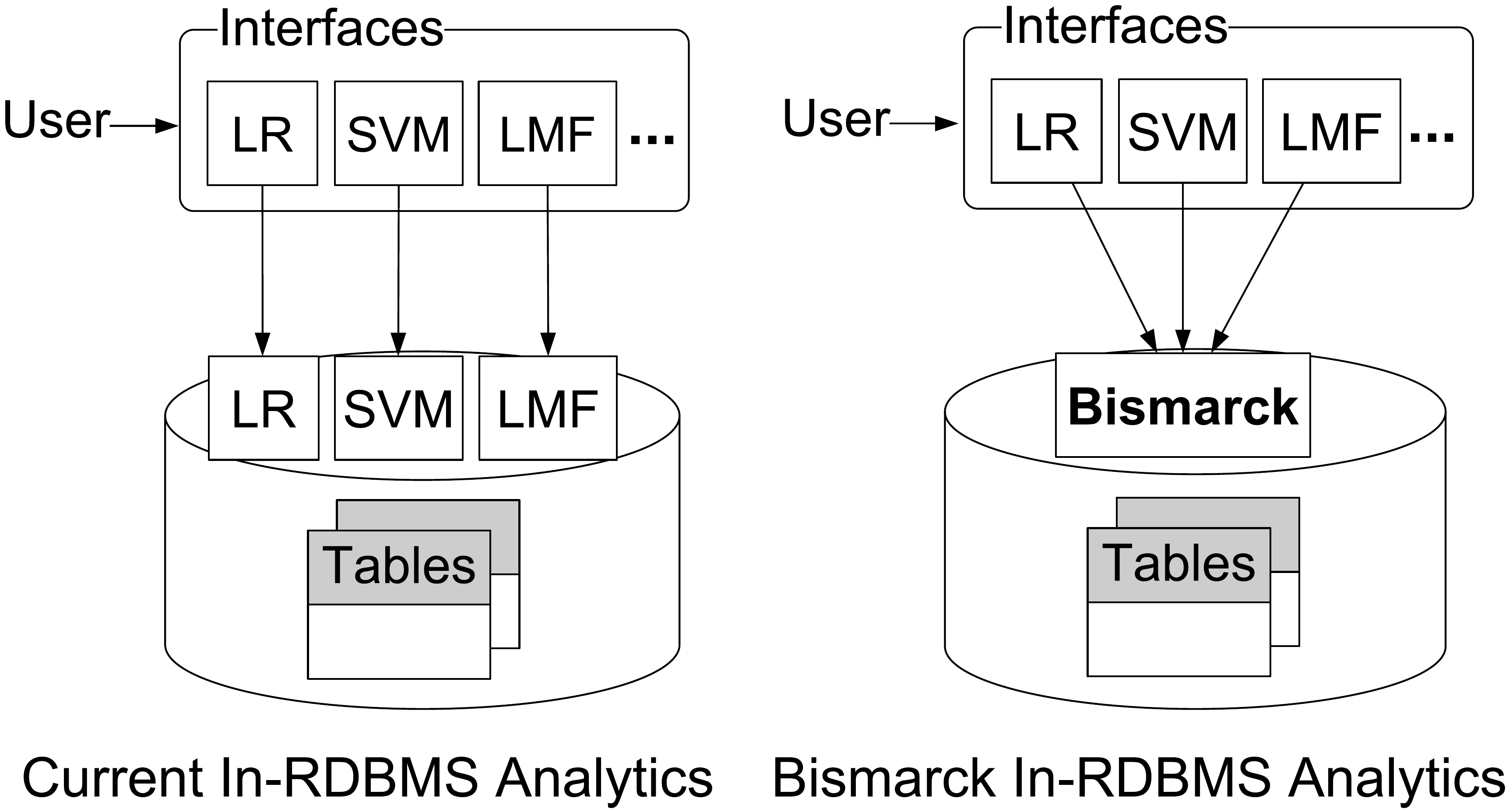} 
 }
&
}
\fullversion{
 \includegraphics[width=4.5in]{codepath}\\
 \vspace{0.2in}
}
\renewcommand\arraystretch{1.5}
\begin{tabular}{|l||c|}
\hline
Analytics Task & Objective \\
\hline
\hline
Logistic Regression (LR) & $\sum_{i} \log (1 + exp (- y_i w^{T} x_i)) + \mu\|\vec{w}\|_1$\\
\hline
Classification (SVM)& $\sum_{i} (1 - y_iw^{T}x_i)_{+} + \mu\|\vec{w}\|_1$\\
\hline
Recommendation (LMF) & $\sum_{(i,j)\in\Omega} (L_i^T R_j - M_{ij})^2 + \mu\|L,R\|_F^2$\\
\hline
Labeling (CRF)~\cite{wallach:crf} & $\sum_{k} \left[ \sum_{j} w_j F_j(y_k, x_k) - \log Z(x_k) \right]$ \\
\hline
Kalman Filters &$\sum_{t=1}^T ||Cw_t-f(y_t)||^2_2 + ||w_t-Aw_{t-1}||^2_2$ \\
\hline
Portfolio Optimization & $p^T w + w^T \Sigma w ~~\mbox{s.t.}~~w\in\Delta$ \\
\hline
\end{tabular}
\submissionversion{
\end{tabular}
}
\caption{\name in an RDBMS: (A) In contrast to existing in-RDBMS
  analytics tools that have separate code paths for different
  analytics tasks, \name provides a single framework to implement
  them, while possibly retaining similar interfaces.  (B) Example
  tasks handled by \name. In Logistic Regression and Classification,
  we minimize the error of a predictor plus a regularization term.  In
  Recommendation, we find a low-rank approximation to a matrix $M$
  which is only observed on a sparse sampling of its entries.  This
  problem is not convex, but it can still be solved via IGD. In
  Labeling with Conditional Random Fields, we maximize the weights
  associated with features $(F_j)$ in the text to predict the
  labels. In Kalman Filters, we fit noisy time series data. In
  quantitative finance, portfolios are optimized balancing risk
  ($p^Tw$) with expected returns ($w^T\Sigma w$); the allocations must
  lie in a simplex, $\Delta$, i.e., $\Delta = \{w \in \mathbb{R}^{n}
  \mid \sum_{i=1}^{n} w_i = 1\}$ and $w_i \geq 0$ for $i=1,\dots,n$.}
\label{fig:objs:grads}
\end{figure*}

The mathematical programming literature is filled with algorithms to
solve convex programming problems. Our first goal is to find an
algorithm in that literature whose data access properties are amenable
to implementation inside an RDBMS. We observe that a classical
algorithm from the mathematical programming cannon, called {\em
  incremental gradient descent} (IGD), has a data-access pattern that
is essentially identical to the data access pattern of any SQL
aggregation function, e.g., an SQL AVG. As we explain in
Section~\ref{sec:prelim}, IGD can be viewed as examining the data one
tuple at time and then performing a (non-commutative) aggregation of
the results. Our first contribution is an architecture that leverages
this observation: {\em we show that we can implement these methods
  using the user-defined aggregate features that are available inside
  every major RDBMS}. To support our point, we implement our
architecture over PostgreSQL and two commercial database systems. In
turn, this allows us to implement all convex data analysis techniques
that are available in current RDBMSes -- and many next generation
techniques (see Figure~\ref{fig:objs:grads}). The code to add in a new
model can be as little as ten lines of C code, e.g., for logistic
regression.\footnote{\scriptsize{Not all data analysis problems are
  convex. Notable exceptions are Apriori
  \cite{DBLP:conf/vldb/AgrawalS94} and graph mining algorithms.}}

As with any generic architectural abstraction, a key question is to understand 
how much performance overhead our approach would incur. In the two commercial
systems that we investigate, we show that compared to a strawman
user-defined aggregate that computes no value, our approach has
between 5\% (for simple tasks like regression) to 100\% overhead (for
complex tasks like matrix factorization). What is perhaps more
surprising is that our approach is often much faster than existing
in-database analytic tools from commercial vendors: 
our prototype implementations are in many
cases $2-4$x faster than existing approaches for simple tasks -- and
for some newly added tasks such as matrix factorization, orders of
magnitude faster. 

A second benefit of a unified in-database architecture is that we can
study the factors that impact performance and optimize them in a way
that applies across several analytics tasks. Our preliminary
investigation revealed many such optimization opportunities including
data layout, compression, data ordering, and parallelism. Here, we
focus on two such factors that we discovered were important in our
initial prototype: {\em data clustering}, i.e., how the data is
ordered on-disk, and {\em parallelism} on a single-node multicore
system.

Although IGD will converge to an optimal solution on convex
programming problems no matter how the underlying data is ordered,
empirically some orders allow us to terminate more quickly than
others.  We observe that inside an RDBMS, data is often clustered for
reasons unrelated to the analysis task (e.g., to support efficient
query performance), and running IGD through the data in the order that
is stored on disk can lead to considerable degradation in performance.
With this in mind, we describe a theoretical example that
characterizes some ``bad'' orders for IGDs and shows that they are
indeed likely inside an RDBMS. For example, if one clusters the data
for a classification task such that all of the positive examples come
before the negative examples, the resulting convergence rate may be
much slower than if the data were randomly ordered, i.e., to reach the
same distance to the optimal solution, more passes over the data are
needed if the data is examined by IGD in the clustered order versus a
random order.  Our second technical contribution is to describe the
clustering phenomenon theoretically, and use this insight to develop a
simple approach to combat this.  A common approach in machine learning
randomly permutes the data with each pass.  However, such random
shuffling may incur substantial computational overhead.  Our method
obviates this overhead by shuffling the data only once before the
first pass. We implement and benchmark this approach on all three
RDBMSes that we study: empirically, we find that across a broad range
of models, while shuffling once has a slightly slower convergence rate
than shuffling on each pass, the lack of expensive reshuffling allows
us to simply run more epochs in the same amount of time. Thus,
shuffling once has better overall performance than shuffling always.

We then study how to parallelize IGD in an RDBMS. We first observe
that recent work in the machine learning community allows us to
parallelize IGD~\cite{DBLP:conf/icdm/ZhuCWZC09} in a way that
leverages the standard user-defined aggregation features available in
every RDBMS to do shared-nothing parallelism. We leverage this parallelization
feature in a commercial database and show that we can get almost
linear speed-ups.  However, recent results in the machine learning
community have shown that these approaches may yield suboptimal
runtime performance compared to approaches that exploit shared-memory
parallelism~\cite{hogwild, DBLP:journals/jmlr/LangfordLZ09}.
This motivates us to adapt approaches that exploit shared memory for
use inside an RDBMS.  We focus on single-node multicore parallelism
where shared memory is available. Although not in the textbook
description of an RDBMS, all three RDBMSes we inspected allow us to
allocate and manage some shared memory (even providing interfaces to
help manage the necessary data structures). We show that the
shared-memory version converges faster than the shared-nothing
version.

In some cases, a single shuffle of the data may be too expensive
(e.g., for data sets that do not fit in available memory).  To cope
with such large data sets, users often perform a subsampling of the
data (e.g., using a {\em reservoir
  sample}~\cite{DBLP:journals/toms/Vitter85}). Subsampling is not
always desirable, as it introduces an additional error (increasing the
variance of the estimate). Thus, for such large data sets, we would
like to avoid the costly shuffle of the data to achieve better
performance than subsampling. Our final technical contribution
combines the parallelization scheme and \textit{reservoir sampling} to
get our highest performance results for datasets that do not fit in
available RAM. On simple tasks like logistic regression, we are 4X
faster than state-of-the-art in-RDBMS tools.  On more complex tasks
like matrix factorization, these approaches allow us to converge in a
few hours, while existing tools do not finish even after several
days.

In summary, our work makes the following contributions:
\begin{CompactItemize}
\item We describe a novel unified architecture, \name, for integrating many
  data analytics tasks formulated as Incremental Gradient Descent into
  an RDBMS using features available in almost every commercial and
  open-source system. We give evidence that our architecture is widely
  applicable by implementing \name in three RDBMS engines:
  PostgreSQL and two commercial engines.

\item We study the effect of data clustering on performance. We
  identify a theoretical example that shows that bad orderings not
  typically considered in machine learning do occur in databases and
  we develop a novel strategy to improve performance. 

\item We study how to adapt existing approaches to make \name run in
  parallel. We verify that this allows us to achieve large speed-ups
  on a wide range of tasks using features in existing RDBMSes. We
  combine our solution for clustering with the above parallelization schemes
  to attack the problem of bad data ordering.
\end{CompactItemize}

We validate our work by implementing \name on three RDBMS
engines: PostgreSQL, and two commercial engines, \ADB and
\BDB. We perform an extensive experimental validation. We see
that we are competitive, and often better than state-of-the-art
in-database tools for standard tasks like regression and
classification. We also show that for next generation tasks like
conditional random fields, we have competitive performance against
state-of-the-art special-purpose tools.

\paragraph*{Related Work} Every major database vendor has data mining tools
associated with their RDBMS offering. Recently, there has been an
escalating arms race to add sophisticated analytics into the RDBMS
with each iteration bringing more sophisticated tools into the
RDBMS. So far, this arms race has centered around bringing individual
statistical data mining techniques into an RDBMS, notably Support
Vector Machines~\cite{DBLP:conf/vldb/MilenovaYC05}, Monte Carlo
sampling~\cite{DBLP:conf/sigmod/JampaniXWPJH08,
  DBLP:journals/pvldb/WickMM10b}, Conditional Random
Fields~\cite{DBLP:journals/pvldb/WangFGH10,DBLP:conf/vldb/GuptaS06},
and Graphical
Models~\cite{DBLP:journals/pvldb/SenDG08,DBLP:journals/pvldb/WangMGH08}. Our
effort is inspired by these approaches, but the goal of this work is
to understand the extent to which we can handle these analytics tasks
with a single unified architecture. Of these approaches, 
MCDB~\cite{DBLP:conf/sigmod/JampaniXWPJH08} and Wick et
al.~\cite{DBLP:journals/pvldb/WickMM10b} are the most related in that
they propose a single unified interface for uncertain data based on
sampling and graphical models respectively. In contrast, we consider
data analytics techniques that are modeled as convex programming
problems.

A related (but orthogonal issue) is how statistical models should be
integrated into the RDBMS to facilitate ease of use, notably {\em
  model-based views} pioneered in
MauveDB~\cite{DBLP:conf/sigmod/DeshpandeM06}. The idea is to give
users a unified abstraction that hides from the user (but not the tool
developer) the details of statistical processing. In contrast, our
goal is a lower level abstraction: we want to unify at the
implementation of many different data analysis tasks.

Using incremental gradient algorithms for convex programming problems
is a classical idea, going back to the seminal work in the 1950s of
Robbins and Monro~\cite{RobbinsMonro51}. Recent years have seen a
resurgence of interest in these algorithms due to their ability to
tolerate noise, converge rapidly, and achieve high runtime
performance. In fact, sometimes an IGD method can converge before
examining all of the data; in contrast, a traditional gradient method
would need to touch all of the data items to take even a single
step. These properties have made IGD an algorithm of choice in the Web
community. Notable implementations include Vowpal Wabbit at
Yahoo!~\cite{VowpalWabbit}, and in large-scale
learning~\cite{DBLP:conf/nips/BottouB07}. IGD has also been employed
for specific algorithms, notably Gemulla et al recently used it for
matrix factorization~\cite{DBLP:conf/kdd/GemullaNHS11}. What
distinguishes our work is that we have observed that IGD forms the
basis of a systems abstraction that is well suited for in-RDBMS
processing. As a result, our technical focus is on optimizations that
are implementable in an RDBMS and span many different models.

Our techniques to study the impact of sorting is inspired by the work
of Bottou and LeCun~\cite{DBLP:conf/nips/BottouC03}, who
empirically studied the related problem of different sampling
strategies for stochastic gradient algorithms. There has been a good
deal of work in the machine learning community to create several
clever parallelization schemes for
IGD~\cite{BertsekasNLP,DBLP:journals/jmlr/LangfordLZ09,Dekel11,Zinkevich10,Duchi10}. Our
work builds on this work to study those methods that are ideally
suited for an RDBMS. For convex programming problems, we find that the
model averaging techniques of Zinkevich et al~\cite{Zinkevich10} fit
well with user-defined aggregates. Recently, work on using shared
memory {\em without locking} has been shown to converge more rapidly
in some settings~\cite{hogwild}. We borrow from both
approaches.

Finally, the area of convex programming problems is a hot topic in
data analysis~\cite{boyd:cvx, BertsekasNLP}, e.g., the support vector
machine~\cite{Mangasarian65}, Lasso~\cite{Tibshirani96}, and logistic
regression~\cite{Wahba95} were all designed and analyzed in a convex
programming framework. Convex analysis also plays a pivotal role in
approximation algorithms, e.g., the celebrated MAX-CUT
relaxation~\cite{DBLP:journals/jcss/GoemansW04} shows that the optimal
approximation to this classical \textsf{NP}-hard problem is achieved by
solving a convex program. In fact a recent result in the Theory community
shows that there is reason to believe that almost all combinatorial
optimization problems have optimal approximations given by solving
convex programs~\cite{DBLP:conf/stoc/Raghavendra08}. Thus, we argue
that these techniques may enable a number of sophisticated data
processing algorithms in the RDBMS.

\paragraph*{Outline} The rest of the paper is organized as follows:
In Section~\ref{sec:prelim}, we explain how \name interacts with the
RDBMS, and give the necessary mathematical programming background on
gradient methods. In Section~\ref{sec:arch}, we discuss the
architecture of \name, and how data ordering and parallelism impact
performance. In Section~\ref{sec:experiments}, we validate that \name
is able to integrate analytics techniques into an RDBMS with low
overhead and high performance.


\section{Preliminaries}
\label{sec:prelim}

We start with a description of how \name fits into an RDBMS, and then
give a simple example of how an end-user interacts with \name in an
RDBMS.  We then discuss the necessary mathematical programming background on gradient methods.

\subsection{\name in an RDBMS}

We start by contrasting the high level architecture of most existing
in-RDBMS analytics tools with how \name integrates analytics into
an RDBMS, and explain how \name is largely orthogonal to the 
\textit{end-user interfaces}.
Existing tools like MADlib \cite{DBLP:journals/pvldb/CohenDDHW09},
Oracle Data Mining \cite{oracle-dm:web}, and 
Microsoft SQL Server Data Mining \cite{ms-dm:web}
provide SQL-like interfaces for the end-user to specify tasks 
like Logistic Regression, Support Vector Machine, etc. Declarative 
interface-level abstractions like \textit{model-based views} \cite{DBLP:conf/sigmod/DeshpandeM06} 
help in creating such user-friendly interfaces. 
However, the underlying implementations of these tasks do not have a 
unified architecture, increasing the overhead for the developer.
In contrast, \name provides a single architectural abstraction
for the developer to unify the \textit{in-RDBMS implementations} of these
analytics techniques, as illustrated in Figure \ref{fig:objs:grads}.
Thus, \name is orthogonal to the end-user interface, and the developer 
has the freedom to provide any existing or new interfaces. In fact, in our
implementation of \name in each RDBMS, \name's user-interface mimics
the interface of that RDBMS' native analytics tool.

For example, consider the
interface provided by the open-source MADlib
\cite{DBLP:journals/pvldb/CohenDDHW09} used over PostgreSQL and
Greenplum databases.  Consider the task of classifying papers using a support
vector machine (SVM).  The data is in a table {\tt LabeledPapers(id,
  vec, label)}, where {\tt id} is the key, {\tt vec} is the feature
values (say as an array of floats) and {\tt label} is the class label.
In MADlib, the user can train an SVM model by simply issuing a SQL
query with some pre-defined functions that take in the data table
details, parameters for the model, etc.
\cite{DBLP:journals/pvldb/CohenDDHW09} In \name, we mimic this
familiar interface for users to do in-RDBMS analytics. For example,
the query (similar to MADlib's) to train an SVM is as follows:\\

\submissionversion{{\tt SELECT SVMTrain (`myModel', `LabeledPapers',

\hspace{1.1 in}`vec', `label');}\\}

\fullversion{{\tt SELECT SVMTrain (`myModel', `LabeledPapers',
`vec', `label');}\\}

\noindent
SVMTrain is a function that passes the user inputs to \name, which
then performs the gradient computations for SVM and returns the model.
The model, which is basically a vector of coefficients for an SVM, is
then persisted as a user table {\tt `myModel'}.  The model can be
applied to new unlabeled data to make predictions by using a similar SQL
query.

\subsection{Background: Gradient Methods}
\label{sec:grad}

We provide a brief introduction to gradient methods. For a thorough
introduction to gradient methods and their projected, incremental
variants, we direct the interested reader to the many
surveys of the subject~\cite{Nemirovski09,BertsekasSurvey}.  We focus
on a particular class of problems that have {\em linearly separable}
objective functions. Formally, our goal is to find a vector $w \in
\mathbb{R}^{d}$ for some $d \geq 1$ that minimizes the following
objective:\footnote{\scriptsize{In Appendix~\ref{a:grads}, we
    generalize to include constraints via proximal point methods. One
    can also generalize to both matrix valued $w$ and
    non-differentiable functions~\cite{Rockafellar1996Convex}.}}
\begin{equation}
 \min_{w \in \mathbb{R}^{d}} \quad \sum_{i=1}^{N} f(w,z_i) + P(w)
\label{eq:obj}
\end{equation}
\noindent
Here, the objective function decomposes into a sum of functions
$f(w,z_i)$ for $i=1,\dots,N$ where each $z_i$ is an item of (training)
data. In \name, the $z_i$ are represented by tuples in the database,
e.g., a pair (paper,area) for paper classification. We abbreviate
$f(w,z_i) = f_i(w)$. For example, in SVM classification, the function
$f_i(w)$ could be the hinge loss of the model $w$ on the $i$th data
element and $P(w)$ enforces the smoothness of the classifier
(preventing overfitting). Eq.~\ref{eq:obj} is general:
Figure~\ref{fig:objs:grads}(B) gives an incomplete list of examples that
can be handled by \name.

A gradient is a generalization of a derivative that tells us if the
function is increasing or decreasing as we move in a particular
direction. Formally, a {\em gradient} of a function $h :
\mathbb{R}^{d} \to \mathbb{R}$ is a function $\nabla h :
\mathbb{R}^{d} \to \mathbb{R}^{d}$ such that $(\nabla h(w))_{i} =
\frac{\partial}{\partial w_{i}} h(w)$~\cite{boyd:cvx}. Linearity of
the gradient implies the equation:
\[ \nabla \sum_{i=1}^{N} f_i(w) = \sum_{i=1}^{N} \nabla f_i(w) \,.\]
For our purpose, the importance of this equation is that to compute
the gradient of the objective function, we can compute the gradient of
each $f_i$ individually.

\emph{Gradient methods} are algorithms that solve
(\ref{eq:obj}). These methods are defined by an iterative rule that
describes how one produces the $(k+1)$-st iterate, $w^{(k+1)}$, given
the previous iterate, $w^{(k)}$.  For simplicity, we assume that
$P=0$. Then, we are minimizing a function $f(w) = \sum_{i=1}^{N}
f_i(w)$, our goal is to produce a new point $w^{(k+1)}$ where
$f(w^{(k)}) > f(w^{(k+1)})$. In 1-D, we need to move in the direction
opposite the derivative (gradient). A gradient method is defined by
the rule:
\[ w^{(k+1)} = w^{(k)} - \alpha_k \nabla f(w^{(k)}) \]
here $\alpha_k \geq 0$ is a positive parameter called {\em step-size}
that determines how far to follow the current search
direction. Typically, $\alpha_k \to 0$ as $k \to \infty$.

The twist for {\em incremental} gradient methods is to approximate the
full gradient using a single terms of the sum. That is, let $\eta(k)
\in \{1,\ldots,N\}$, chosen at iteration $k$. Intuitively, we
approximate the gradient $\nabla f(w)$ with $\nabla
f_{\eta(k)}(w)$.\footnote{\scriptsize{Observe that minimizing $f$ and $g(w) =
  \frac{1}{N}f(w)$, means correcting by the factor $N$ is not
  necessary and not done by convention.}} Then,
\begin{equation}
 w^{(k+1)} = w^{(k)} - \alpha_k \nabla f_{\eta(k)}(w^{(k)})
\label{eq:rule}
\end{equation}

\noindent
This is a key connection: each $f_i$ can be represented as a single
tuple. We illustrate this rule with a simple example:

\begin{example}
Consider a simple least-squares problem with $2n$ ($n \geq 1$) data
points $(x_1,y_1),$ $\dots,$ $(x_{2n}, y_{2n})$.  The feature values
are $x_i = 1$ for $i=1,\dots, 2n$ and the labels are $y_i = 1$ for $i
\leq n$, and $y_i = -1$, otherwise. The resulting mathematical
programming problem is:
  \[ \min_{w} \quad \frac{1}{2} \sum_{i=1}^{2n} (wx_i - y_i)^2 \] 
Since $x_i = 1$ for all $i$, the optimal solution to the problem is the mean $w=0$, but
we choose this to illustrate the mechanics of the method. We begin
with some point $w^{(0)}$ chosen arbitrarily. We choose $i \in
\{1,\dots,2n\}$ at random. Fix some $\alpha \geq 0$ and for $k \geq
0$, set $\alpha_k = \alpha$ for simplicity. Then, our approximation to
the gradient is $\nabla f_i(w^{(0)}) = (w^{(0)} - y_i)$. And so, our
first step is:
\[ w^{(1)} = w^{(0)} - \alpha (w^{(0)} - y_i) \]
We then repeat the process with $w^{(2)}$, etc. One can check that
after $k+1$ steps, we will have:
\[ w^{(k+1)} = (1-\alpha)^{k+1} w_0 + \alpha \sum_{j=0}^k (1-\alpha)^{k-i} y_{\eta(j)} \]
Since the expectation of $y_{\eta(j)}$ equals $0$, we can see that we
converge exponentially quickly to $0$ under this scheme -- even before
we see all $2n$ points. This serves to illustrate why an IGD scheme
may converge much faster than traditional gradient methods, where one
must touch every data item at least once just to compute the first
step.
\label{ex:mean}
\end{example}

Remarkably, when both the functions $\sum_{i=1}^n f_i(w)$ and $P(w)$
are both convex, the incremental gradient method is guaranteed to
converge to a globally optimal solution~\cite{Nemirovski09} at known
rates. Also, IGD converges (perhaps at a slower rate) even if
$\eta(k)$ is a sequence in a fixed, arbitrary
order~\cite{Luo91,Mangasarian94,Luo94,Bertsekas97,Tseng98}. We explore
this issue in more detail in Example~\ref{ex:txca}.

\section{Bismarck Architecture}
\label{sec:arch}
We first describe the high-level architecture of \name, and then explain how we 
implement IGD in an RDBMS. Then, we drill down into two aspects 
of our architecture that impact performance - {\it data ordering} and {\it parallelism}.

\begin{figure}
\centering
 \includegraphics[width=2.6in]{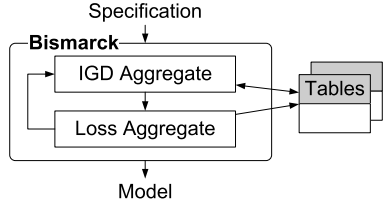} 
\caption{High-level Architecture of \name.}
\label{fig:higharch}
\end{figure}

\subsection{High-Level Architecture}
\label{sec:higharch}
\eat{
\chris{This section should be at a systems level (no user
  mentioned). State clearly the input: a developer needs to write
  code. The setup here is so that later you can precisely describe
  shuffles, etc. Also, rewrite the description to say ``HERE IS WHAT A
  USER DEFINED AGGREGATE DOES'' and then explain it in that way. So
  that you can later make it clear that the parallelization is for
  free in 3.3}
}

The high-level architecture of \name is presented in Figure
\ref{fig:higharch}.  \name takes in the specifications for an
analytics task (e.g., data details, parameters, etc.)  and runs the
task using Incremental Gradient Descent (IGD). As explained before,
IGD allows us to solve a number of analytics tasks in one unified
way. The main component of \name is the in-RDBMS implementation of IGD
with a data access pattern similar to a SQL aggregate query. For this
purpose, we leverage the mechanism of User-Defined Aggregate (UDA), a
standard feature available in almost all RDBMSes
\cite{uda-postgres:web,uda-oracle:web,uda-sqlserver:web}. The UDA
mechanism is used to run the IGD computation, but also to test for
convergence and compute information, e.g., error rates. \name also
needs to provide a simple iteration to test for convergence. We will
explain more about these two aspects shortly, but first we describe
the architecture of a UDA, and how we can handle IGD in this
framework.

\begin{figure}
\centering
 \includegraphics[width=2.6in]{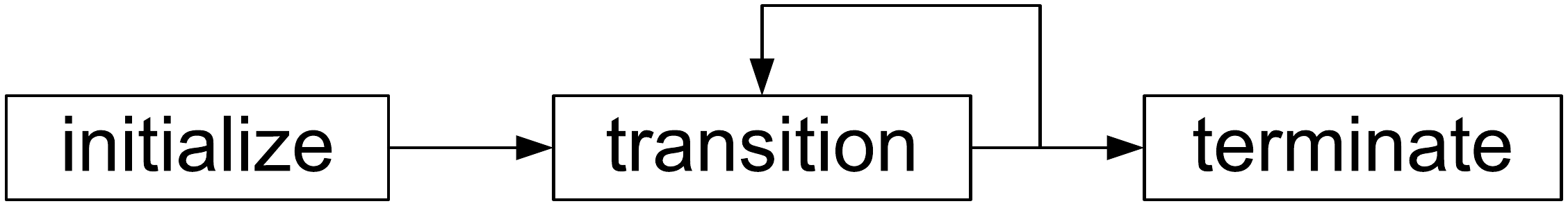}
\caption{The Standard Three Phases of a UDA.}
\label{fig:arch}
\end{figure}

\paragraph*{IGD as a User-Defined Aggregate}
As shown in Figure \ref{fig:arch}, a developer implements a UDA by
writing three standard functions: {\tt initialize(state)}, {\tt
  transition(state, data)} and {\tt terminate(state)}.  Almost all
RDBMSes provide the abstraction of a UDA, albeit with different names
or interfaces for these three steps, e.g., PostgreSQL names them
`initcond', `sfunc' and `finalfunc' \cite{uda-postgres:web}.

The {\tt state} is basically the context of aggregation (e.g., the
running total and count for an {\tt AVG} query). The {\tt data} is a
tuple in the table.  In our case, the {\tt state} is essentially the
\textit{model} (e.g., the coefficients of a logistic regressor) and
perhaps some meta data (e.g., number of gradient steps taken). In our
current implementation, we assume that the {\tt state} fits in memory
(models are typically orders of magnitude smaller than the data, which
is not required to fit in memory). The {\tt data} is again an example
from the data table, which includes the attribute values and the label
(for supervised schemes). We now explain the role of each function:

\begin{itemize}
 \item The {\tt initialize(state)} function initializes the model with user-given values
	(e.g., a vector of zeros), or a model returned by a previous run.
 \item In {\tt transition(state, data)}, we first compute the
        (incremental) gradient value of the objective function on the
        given {\tt data} example, and then update the current model
        (Equation \ref{eq:rule} from Section \ref{sec:grad}).  This
        function is where one puts the logic of the the various
        analytics techniques -- each technique has its own objective
        function and gradient (Figure \ref{fig:objs:grads}(B)). Thus,
        the main differences in the implementations of the various
        analytics techniques occur mainly in a few lines of code
        within this function, while the rest of our architecture is
        reused across techniques. 
        Figure \ref{fig:code} illustrates the claim with
        actual code snippets for two tasks (LR and SVM).        
        This simplifies the development of
        sophisticated in-database analytics, in contrast to existing
        systems that usually have different code paths for different
        techniques (Figure \ref{fig:objs:grads}(A)).
 \item In {\tt terminate(state)}, we finish the gradient computations and return the model, possibly 
	persisting it.
\end{itemize}

\begin{SaveVerbatim}{LRGrad}
  LR_Transition(ModelCoef *w, Example e) { ...
    wx = Dot_Product(w, e.x);
    sig = Sigmoid(-wx * e.y);
    c = stepsize * e.y * sig;
    Scale_And_Add(w, e.x, c); ... }
\end{SaveVerbatim}

\begin{SaveVerbatim}{SVMGrad}
      SVM_Transition(ModelCoef *w, Example e) { ...
        wx = Dot_Product(w, e.x);
        c = stepsize * e.y;
        if(1 - wx * e.y > 0) {
          Scale_And_Add(w, e.x, c); } ... }
\end{SaveVerbatim}

\begin{figure*}[htbp]
	\fbox{
		\centering
		\begin{minipage}{0.975\textwidth}
			\fullversion{\footnotesize}
			\BUseVerbatim{LRGrad}
			\BUseVerbatim{SVMGrad}
		\end{minipage}
	}
	\caption{Snippets of the C-code implementations of the \texttt{transition} step
	for Logistic Regression (LR) and Support Vector Machine (SVM). 
	Here, \texttt{w} is the coefficient vector, and \texttt{e} 
	is a training example with feature vector \texttt{x} and 
	label \texttt{y}. \texttt{Scale\_And\_Add} updates \texttt{w} 
	by adding to it \texttt{x} multiplied by the scalar \texttt{c}.
	Note the minimal differences between the two implementations.
	}
	\label{fig:code}
\end{figure*}

A key implementation detail is that \name may reorder the data to
improve the convergence rate of IGD or to sample from the data. This
feature is supported in all major RDBMSes, e.g., in PostgreSQL using the
\texttt{ORDER BY RANDOM()} construct.

\paragraph*{Key Differences: Epochs and Convergence}
A key difference from traditional aggregations, like SUM, AVG, or MAX,
is that to reach the optimal objective function value, IGD may need to
do more than one pass over the dataset. Following the machine learning
literature, we call each pass an \textit{epoch}~\cite{DBLP:conf/nips/BottouC03}. Thus,
the aggregate may need to be executed more than once, with the output
model of one run being input to the next (shown in Figure
\ref{fig:higharch} as a loop).  To determine how many epochs to run,
\name supports an arbitrary Boolean function to be called (which may
itself involve aggregation). This supports both what we observed in
practice as common heuristic convergence tests, e.g., run for a fixed number of
iterations, and more rigorous conditions based on the norm of the
gradient common in machine learning~\cite{DBLP:journals/mp/AnstreicherW09}.

A second difference is that the we may need to compute the actual
value of the objective function (also known as the \textit{loss})
using the model after each epoch. The loss value may be
needed by the stopping condition, e.g., a common convergence
test is based on the relative drop in the loss value. This loss
computation can also be implemented as a UDA (or piggybacked onto the IGD UDA).

\paragraph*{Technical Opportunities} 
A key conceptual benefit of \name's approach is that one can study
{\em generic} performance optimizations (i.e., optimizations that apply
to many analytics techniques) rather than ad hoc, per-technique ones. 
The remainder of the technical sections are
devoted to examining two such generic optimizations. First, the
conventional wisdom is that for IGD to converge more rapidly, each
data point should be sampled in random (without-replacement)
order~\cite{DBLP:conf/nips/BottouC03}. This can be achieved by randomly reordering, or
\textit{shuffling}, the dataset before running the aggregate for
gradient computation at each epoch. The goal of course is to converge
faster in wall-clock time, not per epoch. Thus, we study when the
increased speed in convergence rate per epoch outweighs the additional
cost of reordering the data at each epoch. The second optimization we
describe is how to leverage multicore parallelism to speed-up the IGD
aggregate computation.

\eat{
\paragraph*{Important Technical Details} 
We exploit shared-memory features offered by every
RDBMS to avoid passing the model around in the {\tt transition}
function on each tuple. We retain the above architecture, but as shown
later in Section \ref{sec:experiments}, the shared-memory variant is
more efficient than the vanilla UDA implementation. Furthermore, for
IGD to converge more rapidly (reach optimal value in fewer epochs),
the conventional wisdom is that the gradient computation should be
done on each data point sampled in random (without-replacement)
order~\cite{DBLP:conf/nips/BottouC03}.  This can be achieved by randomly reordering, or
\textit{shuffling}, the dataset before running the aggregate for
gradient computation at each epoch.
}

\subsection{Impact of Data Ordering}
\label{sec:ordering}

\eat{
\chris{This section should explain the problem (notes do this), then
  give CA-TX example to explain in RDBMS ordering issues, then
  describe the three simple solutions. The solutions are super deep so
  it should be short.}
}

\eat{
As mentioned earlier, the IGD literature typically
assumes that the data is seen in random order, which means we have to shuffle the
dataset at each epoch. Shuffling can be implemented using a clause similar to 
{\tt ORDER BY random()} that is possible in most RDBMSes. However, shuffling at each
epoch could be expensive for large datasets, often dominating the runtime. 
In this section, we analyze this data
ordering issue rigorously and explain how we address it to achieve better performance.
}

On convex programming problems, IGD is known to converge to the optimal value
irrespective of how the underlying data is ordered.  But empirically
some data orderings allow us to converge in fewer epochs than others.
However, our experiments suggest that the sensitivity is not as great as one might
think.  In other words, presenting the data in a random
order gets essentially optimal run-time behavior. This begs the
question as to whether we should even reorder the data randomly at
each epoch.  In fact, some machine learning tools do not even bother
to randomly reorder the data.  However, we observe that inside an
RDBMS, data is often clustered for reasons unrelated to the analysis
task (e.g., for efficient join query performance).  For example, the
data for a classification task might be clustered by the class label.
We now analyze this issue by providing a theoretical example that
characterizes pathological orders for IGD. We chose this example to
illustrate the important points with respect to clustering and be as
theoretically simple as possible.

\begin{figure}
\centering
\includegraphics[width=3.1in]{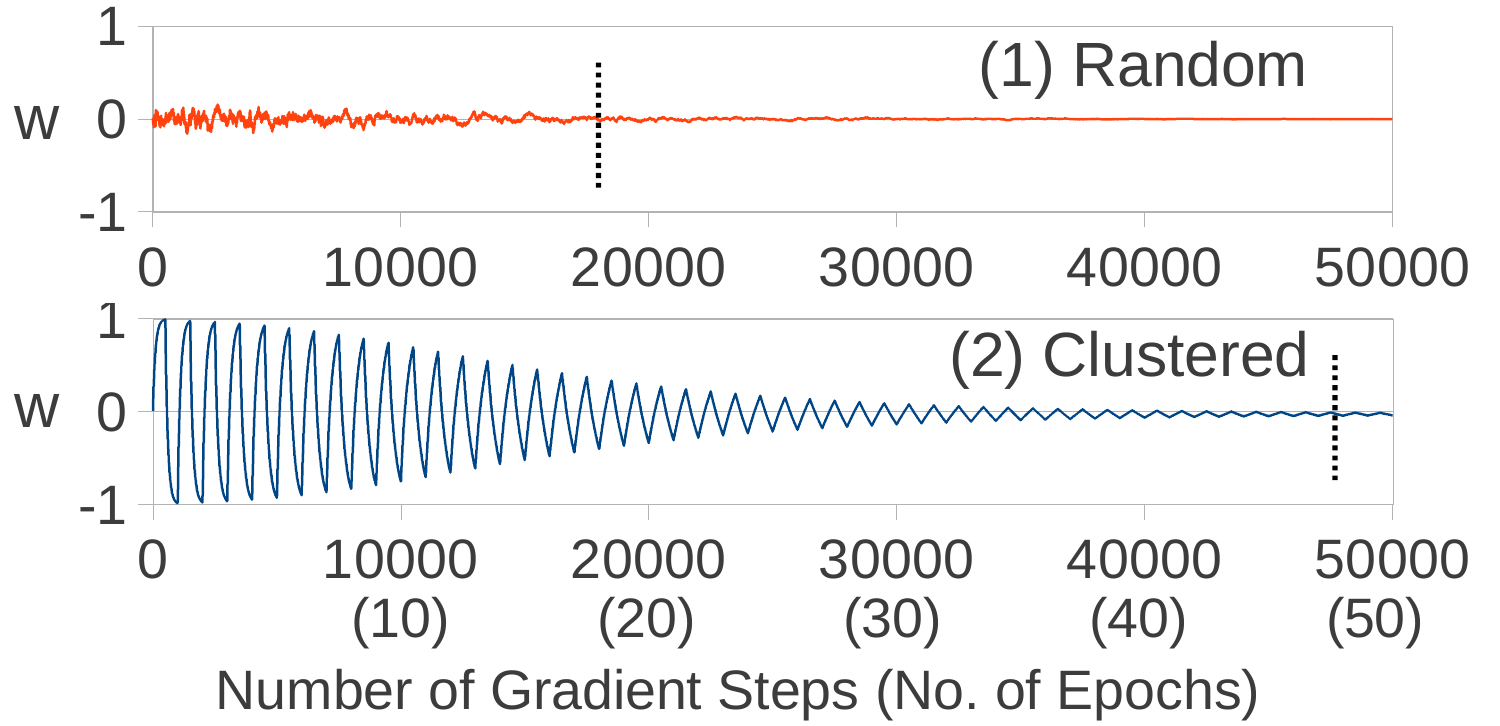}
\caption{1-D CA-TX Example: Plot of $w$ against number of gradient steps on (1) \textit{Random}, and
(2) \textit{Clustered} data orderings for a dataset with 1000 examples (i.e., $n = 500$).
The number of epochs is shown in parentheses on the x-axis. \textit{Random}
takes 18 epochs to converge (convergence defined here as $w^2 < 0.001$), 
while \textit{Clustered} takes 48 epochs.}
\label{fig:pingpong}
\end{figure}

\begin{example}[1-D CA-TX]
\label{ex:txca}
Suppose that our data is clustered geographically, e.g., sales data from 
California, followed by Texas, and the attributes of the sales
in the two states cause the data to be in two different classes. 
With this in mind, recall
Example~\ref{ex:mean}. We are given a simple least-squares problem
with $2n$ ($n \geq 1$) data points $(x_1,y_1),$ $\dots,$ $(x_{2n},
y_{2n})$.  The feature values are $x_i = 1$ for $i=1,\dots, 2n$ and
the labels are $y_i = 1$ for $i \leq n$, and $y_i = -1$,
otherwise. The resulting mathematical programming problem is:
  \[ \min_{w} \quad \frac{1}{2} \sum_{i=1}^{2n} (wx_i - y_i)^2 \] 
  Since $x_i = 1$ for all $i$, the optimal solution is the mean, $w = 0$.  But our goal
  here is to analyze the behavior of IGD on this problem under
  various orders.  Due to this problem's simplicity, we can solve the
  behavior of the resulting dynamical system in closed form under a
  variety of ordering schemes.  Consider two schemes to illustrate our
  point: (1) data points seen are randomly sampled from the dataset,
  and (2) data points seen in ascending index order, $(x_1,y_1),$
  $(x_2,y_2),$ $\dots$.  Scheme (2) simulates operating on data that
  is clustered by class.
  
  Figure~\ref{fig:pingpong} plots the value of $w$ during the course
  of the IGD under the above two sampling schemes (using diminishing
  step-size rule). We see that both approaches do indeed converge to
  the optimal value, but approach (1), which uses random sampling,
  converges more rapidly.  In contrast, in approach (2), $w$
  oscillates between $1$ and $-1$, until converging eventually.
  Intuitively, this is so because the IGD initially takes steps
  influenced by the positive examples, and is later influenced by the
  negative examples (within one epoch).  In other words, convergence
  can be much slower on clustered data.  
  \fullversion{In Appendix \ref{a:txca},}
  \submissionversion{In the extended version of the paper \cite{bismarck-tr},}
  we present calculations to precisely explain this behavior. 
  We conclude
  the example by noting that {\em almost all} permutations of the data
  will behave similar to (1), and not (2). In other words, (2) is a
  pathological ordering, but one which is indeed possible for data
  stored in an RDBMS.
\end{example}

Shuffling the data at each epoch is expensive and incurs a high overhead.
In fact, for simple tasks like LR and SVM, the shuffling time dominates the gradient 
computation time by a factor of 5.
To remove the overhead of shuffling the data at every epoch, while
still avoiding the pathological ordering, we propose a simple solution
-- shuffle the data \textit{only once}.  By randomly reordering the
data once, we avoid the pathological ordering that might be present in
data stored in a database.  We implemented and benchmarked this
approach on all three RDBMSes that we study. As explained later in
Section \ref{sec:exp:ordering}, empirically, we find that shuffling
once suffices across a broad range of models. Shuffling once does have
a slightly lower convergence rate than shuffling always. However,
since we need not shuffle at every epoch, we significantly reduce the
runtime per epoch, which means we can simply run more epochs within
the same wall-clock time so as to reach the optimal value. As we show
later in Section \ref{sec:exp:ordering}, this allows shuffle-once to 
converge faster than shuffle-always (between 2X-6X faster on the 
tasks we studied).

\subsection{Parallelizing Gradient Computations}
\label{sec:parallelism}
\eat{
\chris{I would add in here the ``classical'' approach, and then we
  say the problem is that model averaging empirically may be slower
  has not had great results, so we look at shared-memory. I think we
  should verify this on PSQL}
}

We now study how we can parallelize the IGD aggregate computation to
achieve performance speed-ups on a single-node multicore system.  We
explain two mechanisms for achieving this parallelism -- one based on
standard UDA features, and another based on shared-memory features. We
emphasize that both features are available in almost all RDBMSes.

\paragraph*{Pure UDA Version} 
The UDA infrastructure offered by most RDBMSes (including the commercial
\ADB and \BDB) include an built-in mechanism for
`shared-nothing' parallelism. The RDBMS requires that the developer
provide a function {\tt merge(state, state)}, along with the 3
functions discussed in Section \ref{sec:higharch}.  The {\tt merge}
function specifies how two aggregation contexts that were computed
independently in parallel can be combined.  For example, for an {\tt
  AVG} query, two individual averages with sufficient statistics
(total count) can be combined to obtain a new average.  Generally,
only aggregates that are \textit{commutative} and \textit{algebraic}
can be parallelized in the above
manner~\cite{DBLP:books/aw/AbiteboulHV95}.  Although the IGD is not
commutative, we observe that it is essentially commutative, in that it
eventually converges to the optimal value {\em regardless} the data
order (Section \ref{sec:ordering}). And although the IGD is not
algebraic, recent results from the machine learning community suggest
that one can achieve rapid convergence by averaging models (trained on
different portions of the data)~\cite{Zinkevich10}. Thus, the IGD is
essentially algebraic as well. In turn, this implies that we can use
the parallel UDA approach to achieve near-linear speed-ups on the IGD
aggregate computations.  \eat{However, as we show later in Section
  \ref{sec:experiments}, this model-averaging approach has worse
  convergence (per epoch) than the single-threaded version, though it
  eventually converges to the optimal value on convex programming
  problems.\footnote{\scriptsize{We empirically observed that model-averaging for
  the non-convex problem of LMF converges to the correct value.}}}

\paragraph*{Shared-Memory UDA} 
Shared-memory management is provided by most RDBMSes
\cite{shmem-postgres:web}, and it enables us to implement the IGD
aggregate completely in the user space with no changes needed to the
RDBMS code. This allows us to preserve the 3-function abstraction from
Section \ref{sec:higharch}, and also reuse most of the code from the
UDA-based implementation. 
The model to be learned is maintained in shared 
memory and is concurrently updated by parallel threads operating
on different segments of the data.
Concurrent updates suggest that we need
locking on the shared model. Nevertheless, recent results from the
machine learning community show that IGD can be parallelized in
a shared-memory environment with no locking at
all~\cite{hogwild}. We adopt this technique into \name. 
Light-weight locking schemes often have stronger theoretical properties for
convergence, and so we consider one such scheme called Atomic
Incremental Gradient (AIG) that uses only \textit{CompareAndExchange}
instructions to effectively perform per-component
locking~\cite{hogwild}.

As shown later in Section~\ref{sec:experiments}, we empirically observe that the model-averaging
approach (pure UDA) has a worse convergence rate than the shared-memory UDA,
and so worse overall performance. This led us to consider the shared-memory UDA for \name.

\eat{
\begin{algorithm}
1. Compute gradient $\mathbf{g} \leftarrow G_e(\mathbf{w}) ~~//\mathbf{w} ~is ~the ~model$\\
2. \textbf{for} component $i$ of $e$: ~//$e$ could be sparse\\
3. ~~ $ w_i \leftarrow w_i - \alpha g_i$ ~//$\alpha$ is the step-size\\
\caption{Gradient step (grad) on example $e$}
\label{alg:grad}
\end{algorithm}

} 
\subsection{Avoiding Shuffling Overhead} 

From the CA-TX example in Section~\ref{sec:ordering}, we saw that bad
data orderings can impact convergence, and that shuffling once
suffices in some instances to achieve good convergence rate.  However,
shuffling even once could be expensive for very large datasets. We
verified this on a scalability dataset, and it did not finish shuffling
even in one day. Thus,
we investigate if it is possible to achieve good convergence rate even on
bad data orderings without any shuffling.  A classical technique to
cope with this situation is to {\em subsample} the data using 
\textit{reservoir sampling} (in fact, some vendors do implement subsampling); 
in this technique, given an in-memory buffer size B, we can
obtain a without-replacement sample of size B in just one pass over
the dataset, without shuffling the dataset
\cite{DBLP:journals/toms/Vitter85}.  The main idea of reservoir
sampling is straightforward: suppose that our reservoir (array) can
hold $m$ items and our goal is to sample from $N$ ($\geq m$)
items. Read the first $m$ items and fill the reservoir. Then,
when we read the $k$th additional item ($m+k$ overall), we
randomly select an integer $s$ in $[0,m+k)$. If $s < m$, then we
  put the item at slot $s$; otherwise we drop the item.

Empirically, we observe that the subsampling may have slow
convergence.  Our intuition is that the reservoir discards valuable
data items that could be used to help the model converge faster. To address
this issue, we propose a simple scheme that we call
\textit{multiplexed} reservoir sampling (MRS), which combines the
reservoir sampling idea with the concurrent model updates idea from
Section~\ref{sec:parallelism}.

\eat{
\begin{algorithm}[t]
{\tt initialize}:\\
~~ 1. $\mathbf{signal} \leftarrow true$ ~//token for memory thread to run\\
~~ 2. Fork a memory thread\\
{\tt transition}: ~//given example $e$\\
~~ 1. $d \leftarrow reservoir(D, e)$ ~//draw reservoir sample $d$\\
~~ 2. Do gradient step on $d$ ~//Algorithm \ref{alg:grad}\\
{\tt terminate}:\\
~~ 1. $\mathbf{signal} \leftarrow false$\\
~~ 2. Swap the reservoir buffers $D$ and $S$\\
\caption{I/O Worker's Steps}
\label{alg:ioworker}
\end{algorithm}

\begin{algorithm}[t]
1. \textbf{while} ($\mathbf{signal} = true$):\\
2. ~~ \textbf{for} each $s$ in $S$:\\
3. ~~~~ Do gradient step on $s$ ~//Algorithm \ref{alg:grad}\\
\caption{Memory Worker's Steps}
\label{alg:memworker}
\end{algorithm}
}

\paragraph*{Multiplexed Reservoir Sampling} The multiplexed reservoir
sampling (MRS) idea is to combine, or \textit{multiplex}, gradient
steps over both the reservoir sample and the data that is not put in
the reservoir buffer.  By using the reservoir sample, which is a valuable
without-replacement sample, and the rest of the data in conjunction,
our scheme can achieve faster convergence than subsampling.

\begin{figure}
\centering
\includegraphics[width=3.1in]{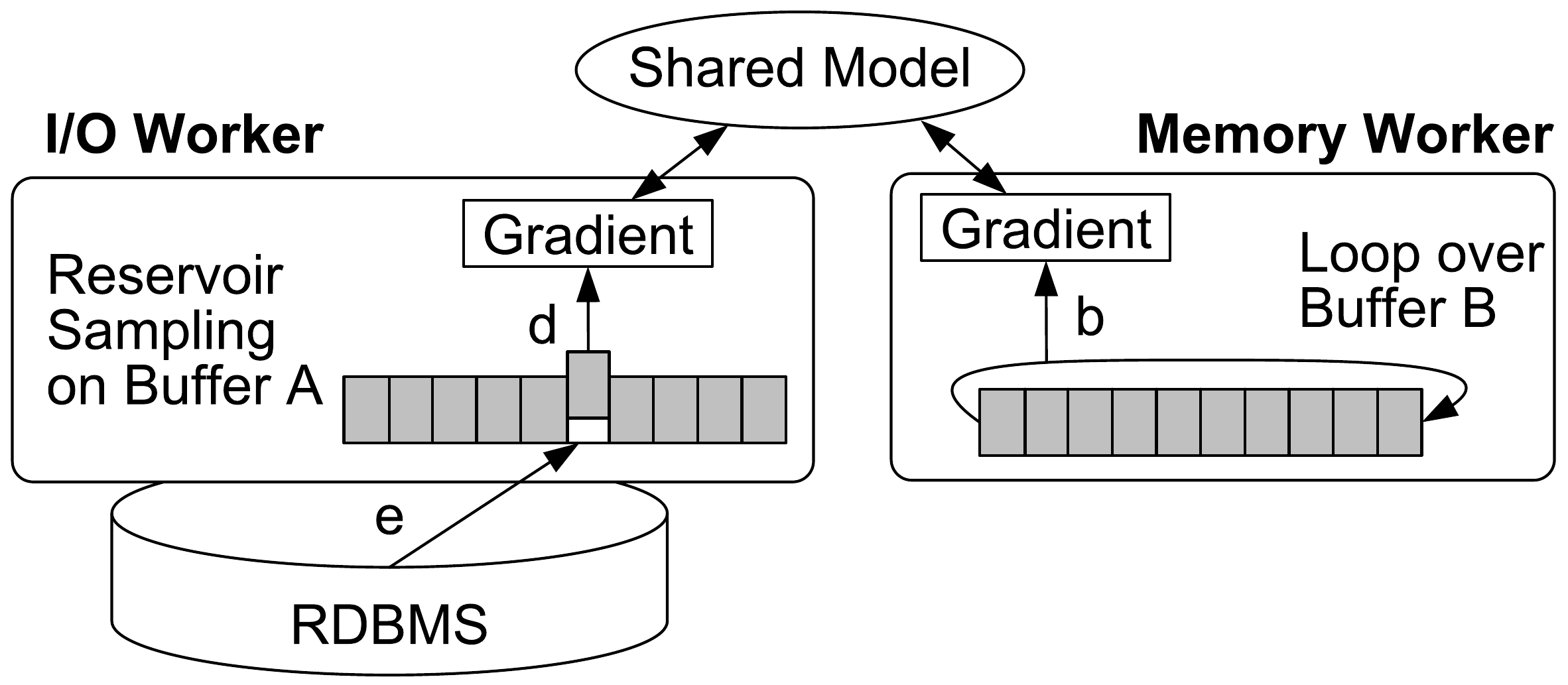}
\caption{Multiplexed Reservoir Sampling (MRS): The I/O Worker reads example tuple $e$ 
from the database, and uses buffer A to do reservoir sampling. The dropped example
$d$ is used for the gradient step, with  updates to a shared model. 
The Memory Worker iterates over buffer B, and performs gradient 
steps on each example $b$ in B concurrently.}
\label{fig:mrs}
\end{figure}

As Figure \ref{fig:mrs} illustrates, in MRS, there are two threads
that update the shared model concurrently, 
called the I/O Worker and the Memory
Worker. The I/O Worker has two tasks: (1) it performs a standard
gradient step (exactly as the previous code), and (2) it places tuples
into a reservoir. Both of these functions are performed within the
previously discussed UDA framework. The Memory Worker takes a buffer as
input, and it loops over that buffer updating the model using the
gradient rule. After the I/O Worker finishes one pass over the data,
the buffers are swapped. That is, the I/O Worker begins filling the buffer
that the Memory Worker is using, while the Memory Worker works on the
buffer that has just been filled by the I/O Worker. The Memory Worker is
signaled by polling a common integer indicating which buffer it should
run over and whether it should continue running. In Section
\ref{sec:experiments}, we show that even with a buffer size that is an
order of magnitude smaller than the dataset, MRS can
achieve better convergence rates than both no-shuffling and
subsampling.



\section{Experiments}
\label{sec:experiments}

We first show that our architecture, \name, incurs little overhead,
in terms of both development effort to add new analytics tasks, and
runtime overhead inside an RDBMS.  We then validate that \name,
implemented over two commercial RDBMSes and PostgreSQL, provides
competitive or better performance than the native analytics tools
offered by these RDBMSes on popular in-database analytics
tasks. Finally, we evaluate how the generic optimizations that we
described in Section \ref{sec:arch} impact \name's performance.

\begin{table}[h]
\centering
\fullversion{\footnotesize}
\begin{tabular}{|c|c|c|c|c|c|}
\hline
Dataset & Dimension & \# Examples & Size\\
\hline
\hline
Forest & 54 & 581k & 77M \\
\hline
DBLife & 41k & 16k & 2.7M\\
\hline
MovieLens & 6k x 4k & 1M & 24M \\
\hline
CoNLL & 7.4M & 9K & 20M \\
\hline
\hline
Classify300M & 50 & 300M & 135G \\
\hline
Matrix5B & 706k x 706k & 5B & 190G \\
\hline
DBLP & 600M & 2.3M & 7.2G\\
\hline
\end{tabular}
\caption{Dataset Statistics. DBLife, CoNLL and DBLP
are in sparse-vector format. MovieLens and Matrix5B are in sparse-matrix format.}
\label{tab:datasets}
\end{table}

\begin{table*}[ht!]
\centering
\fullversion{\footnotesize}
\begin{tabular}{|c|c|c|c||c|c|c|c||c|c|c|c|}
\hline
\multicolumn{4}{|c||}{PostgreSQL} & \multicolumn{4}{|c||}{\ADB} 
& \multicolumn{4}{|c|}{\BDB (8 segments)}\\
\hline
Dataset & \multirow{2}{*}{Tasks} & Run- & Over- &
Dataset & \multirow{2}{*}{Tasks} & Run- & Over- &
Dataset & \multirow{2}{*}{Tasks} & Run- & Over-\\
(\texttt{NULL} time) & & -time & -head &
(\texttt{NULL} time) & & -time & -head &
(\texttt{NULL} time) & & -time & -head \\
\hline
\hline
Forest & LR & 0.57s & 90\% &
Forest & LR & 24.1s & 15.3\% &
Forest & LR & 0.17s & 21.4\%\\
\cline{2-4}
\cline{6-8}
\cline{10-12}
(0.3s) & SVM & 0.56s & 83.3\% &
(20.9s) & SVM & 22.0s & 5.26\% &
(0.14s) & SVM & 0.16s & 14.3\%\\
\hline
DBLife & LR & 0.035s & 192\% &
DBLife & LR & 1.1s & 86.4\% &
DBLife & LR & 0.1 & 17.6\%\\
\cline{2-4}
\cline{6-8}
\cline{10-12}
(0.012s) & SVM & 0.03s & 150\% &
(0.59) & SVM & 0.8s & 35.6\% &
(0.085s) & SVM & 0.096s & 12.9\%\\
\hline
MovieLens & \multirow{2}{*}{LMF} & \multirow{2}{*}{0.86s} & \multirow{2}{*}{244\%} &
MovieLens & \multirow{2}{*}{LMF} & \multirow{2}{*}{45.8s} & \multirow{2}{*}{29.4\%} &
MovieLens & \multirow{2}{*}{LMF} & \multirow{2}{*}{0.32s} & \multirow{2}{*}{100\%}\\
(0.25s) & & & &
(35.4s) & & & &
(0.16s) & & & \\
\hline
\end{tabular}
\caption{{\em Pure UDA} implementation overheads: single-iteration runtime 
of each task implemented in \name against the strawman {\tt NULL} aggregate. 
The parallel database \BDB was run with 8 segments.}
\label{tab:overhead_uda}
\end{table*}

\begin{table*}[ht!]
\centering
\fullversion{\footnotesize}
\begin{tabular}{|c|c|c|c||c|c|c|c||c|c|c|c|}
\hline
\multicolumn{4}{|c||}{PostgreSQL} & \multicolumn{4}{|c||}{\ADB} 
& \multicolumn{4}{|c|}{\BDB (8 segments)}\\
\hline
Dataset & \multirow{2}{*}{Tasks} & Run- & Over- &
Dataset & \multirow{2}{*}{Tasks} & Run- & Over- &
Dataset & \multirow{2}{*}{Tasks} & Run- & Over-\\
(\texttt{NULL} time) & & -time & -head &
(\texttt{NULL} time) & & -time & -head &
(\texttt{NULL} time) & & -time & -head \\
\hline
\hline
Forest & LR & 0.56s & 86.7\% &
Forest & LR & 5.1s & 54.5\% &
Forest & LR & 0.25s & 150\%\\
\cline{2-4}
\cline{6-8}
\cline{10-12}
(0.3s) & SVM & 0.55s & 83.3\% &
(3.3s) & SVM & 4.0s & 21.2\% &
(0.1s) & SVM & 0.21s & 110\%\\
\hline
DBLife & LR & 0.017s & 41.7\% &
DBLife & LR & 0.2s & 81.8\% &
DBLife & LR & 0.045s & 4.6\%\\
\cline{2-4}
\cline{6-8}
\cline{10-12}
(0.012s) & SVM & 0.016s & 33.3\% &
(0.11s) & SVM & 0.3s & 172\% &
(0.043s) & SVM & 0.045s & 4.6\%\\
\hline
MovieLens & \multirow{2}{*}{LMF} & \multirow{2}{*}{0.85s} & \multirow{2}{*}{193\%} &
MovieLens & \multirow{2}{*}{LMF} & \multirow{2}{*}{10.3s} & \multirow{2}{*}{102\%} &
MovieLens & \multirow{2}{*}{LMF} & \multirow{2}{*}{0.26s} & \multirow{2}{*}{160\%}\\
(0.29s) & & & &
(5.1s) & & & &
(0.1s) & & & \\
\hline
\end{tabular}
\caption{{\em Shared-memory UDA} implementation overheads: 
single-iteration runtime of each task implemented in \name 
against the strawman {\tt NULL} aggregate.
The parallel database \BDB was run with 8 segments.}
\label{tab:overhead_shmem}
\end{table*}

\paragraph*{Tasks and Datasets}
We study 4 popular analytics tasks: Logistic Regression (LR), Support
Vector Machine classification (SVM), Low-rank Matrix Factorization
(LMF) and Conditional Random Fields labeling (CRF). We use 4 publicly 
available real-world datasets. For LR and SVM, we
use two datasets -- one dense (Forest, a standard benchmark dataset
from the UCI repository) and one sparse (DBLife, which classifies
papers by research areas).  We binarized these datasets for the
standard binary LR and SVM tasks.  For LMF, we use MovieLens, which is a movie
recommendation dataset, and for CRF, we use the CoNLL dataset, which
is for text chunking. We also perform a scalability study with much
larger datasets -- two synthetic datasets Classify300M (for LR and SVM) and
Matrix5B (for LMF), as well as DBLP (another real-world dataset) for CRF. 
The relevant statistics for all datasets are presented in Table \ref{tab:datasets}.

\paragraph*{Experimental Setup}
All experiments are run on an identical configuration: a dual Xeon X5650 CPUs 
(6 cores each x 2 hyper-threading) machine with 
128GB of RAM and a 1TB dedicated disk. The kernel is Linux 2.6.32-131. 
Each reported runtime is the average of three warm-cache runs. 
Completion time for gradient schemes here means achieving 
0.1\% tolerance in the objective
function value, unless specified otherwise.

\begin{figure*}[ht!]
\centering
\fullversion{\footnotesize}
\submissionversion{
\begin{tabular}{cc}}
\begin{tabular}{|c|c||c|c||c|c||c|c|}
\hline
\multirow{2}{*}{Dataset} & \multirow{2}{*}{Task}
& \multicolumn{2}{|c||}{PostgreSQL} & \multicolumn{2}{|c||}{\ADB} 
& \multicolumn{2}{|c|}{\BDB (8 segments)}\\
\cline{3-8}
& & \name & MADlib & \name & Native & \name & Native\\
\hline
\hline
Forest & LR & 8.0 & 43.5 & 40.2 & 489.0 & 3.7 & 17.0\\
\cline{2-8}
(Dense) & SVM & 7.5 & 140.2 & 32.7 & 66.7 & 3.3 & 19.2\\
\hline
DBLife & LR & 0.8 & N/A & 9.8 & 20.6 & 2.3 & N/A\\
\cline{2-8}
(Sparse) & SVM & 1.2 & N/A & 11.6 & 4.8 & 4.1 & N/A\\
\hline
MovieLens & LMF & 36.0 & 29325.7 & 394.7 & N/A & 11.9 & 17431.3\\
\hline
\end{tabular}
\submissionversion{&}
\fullversion{\vspace{0.2in}\\}
\parbox{1.6in}{
\includegraphics[width=1.6in]{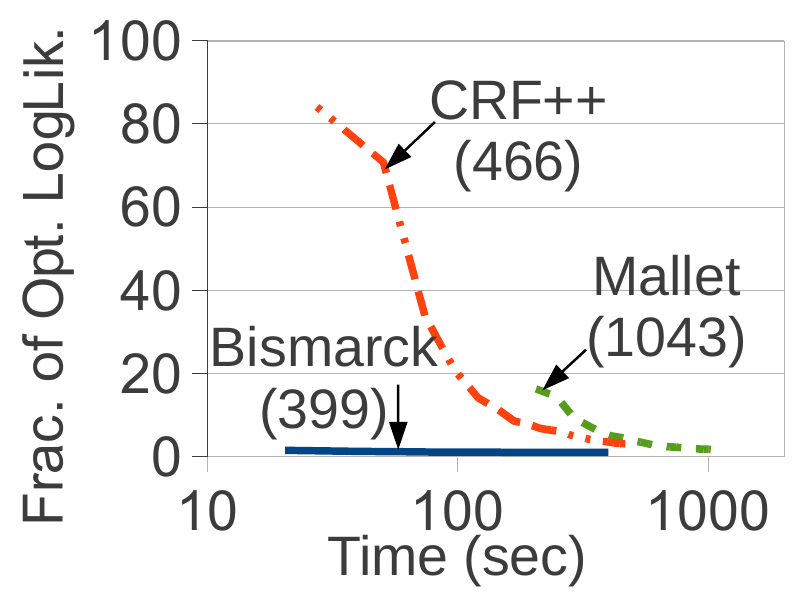}
}
\submissionversion{\end{tabular}}
\caption{Benchmark Comparison: (A) Runtimes (in sec) for convergence
  (0.1\% tolerance) or completion on 3 in-RDBMS analytics tasks. We
  compare \name implemented over each RDBMS against the analytics tool
  native to that RDBMS. N/A means the task is not supported on that
  RDBMS' native tool (B) For the CRF task, we compare \name (over
  PostgreSQL) against custom tools by plotting the objective function
  value against time. Completion times (in sec) are shown in
  parentheses.}
\label{tab:bench}
\end{figure*}

\subsection{Overhead of Our Architecture}
We first validate that \name incurs little development overhead to add
new analytics tasks. We then empirically verify that the runtime
overhead of the tasks in \name is low compared to a strawman
aggregate.

\paragraph*{Development Overhead}
We implemented the 4 analytics tasks in \name over three RDBMSes
(PostgreSQL, commercial \ADB and \BDB). \name enables rapid
addition of a new analytics task since a large fraction of the code is
shared across all the techniques implemented (on a given RDBMS). For
example, starting with an end-to-end implementation of LR in \name (in
C, over PostgreSQL), we need to modify fewer than two dozen lines of
code in order to add the SVM module.\footnote{\scriptsize{Both our code 
and the data used in our experiments are available at: 
http://research.cs.wisc.edu/hazy/victor/bismarck-download/}}
Similarly, we can easily add in a
more sophisticated task like LMF with only five dozen new lines of
code. We believe that this is possible because our unified
architecture based on IGD abstracts out the logic of the various tasks
into a small number of generic functions.  This is in contrast to
existing systems, where there is usually a dedicated code stack for
each task.

\paragraph*{Runtime Overhead}
We next verify that the tasks implemented in \name have low runtime
overhead. To do this, we compared our implementation to a strawman
aggregate that sees the same data, but computes no values. We call
this a NULL aggregate. We run three tasks -- LR, SVM and LMF in \name
over all the 3 RDBMSes, using both the pure UDA infrastructure
(shared-nothing) and the shared-memory variant described in Section
\ref{sec:arch}.  We compare the single-iteration runtime
of each task against the NULL aggregate 
for both implementations of \name over the same datasets.
The results are presented in Tables \ref{tab:overhead_uda} and
\ref{tab:overhead_shmem}.

We see that the overhead compared to the {\tt NULL} aggregate can be
as low as 4.6\%, and is rarely more than 2X runtime for simple tasks
like LR and SVM. The overhead is higher for the more
computation-intensive task LMF, but is still less than 2.5X runtime of
the {\tt NULL} aggregate. We also see that the shared-memory variant
is several times faster than the UDA implementation over \ADB,
since \ADB has extra overheads (e.g., model passing,
serializations, etc.) to run the pure UDA. It was this observation
that prompted us to use the shared-memory UDA to implement \name even
for a single-thread RDBMS.

\subsection{Benchmark Comparison}
\eat{
\begin{itemize}
 \item Goal: Benchmark performance-quality and scalability of Victor against popular ML tools
 \item Hypothesis: Victor's performance is competitive or better on both simple and 
 complex tasks; Victor scales well on all tasks including complex tasks
\end{itemize}
}

We now validate that \name implemented over two commercial RDBMSes and
PostgreSQL provides competitive or better performance than the native
analytics tools offered by these RDBMSes on three existing in-database
analytics tasks -- LR, SVM and LMF.  For the comparison, we use the
shared-memory UDA implementation of \name along with the shuffle-once
approach described in Section \ref{sec:ordering}. For the parallel
version of \name, we use the no-lock shared-memory parallelism
described in Section \ref{sec:parallelism}.

\paragraph*{Competitor Analytics Tools}
We compare \name against three existing in-RDBMS tools -- MADlib (an
open-source collection of in-RDBMS statistical techniques
\cite{DBLP:journals/pvldb/CohenDDHW09}), which is run over PostgreSQL
(single-threaded), and the native analytics tools provided by the two
commercial engines -- \ADB (single-threaded), and the parallel
\BDB (with 8 segments).  We tuned the parameters for each tool,
including \name, on each task based on an extensive search in the
parameter space.  The data was preprocessed appropriately for all
tools.  Some of the tasks we study are not currently supported in the
above tools. In particular, the CRF task is not available in any of
the existing in-RDBMS analytics tools we considered, and so we compare
\name (over PostgreSQL) against the custom tools CRF++ \cite{crfpp}
and Mallet \cite{mallet}.

\paragraph*{Existing In-RDBMS Analytics Tasks}
We first compare the end-to-end runtimes of the various tools on LR,
SVM and LMF. The results are summarized in Figure \ref{tab:bench} (A).
Overall, we see that \name implemented over each RDBMS has competitive
or faster performance on all these tasks against the native tool of
the respective RDBMS.  On simple tasks like LR and SVM, we see that
\name is often several times faster than existing tools.  That is, on
the dense LR task, \name is about 12X faster than \ADB's tool, and
about 5X faster than MADlib over both PostgreSQL and the native tool
in \BDB. In some cases, e.g., \ADB for sparse SVM, \name is
slightly slower due to the function call overheads in \ADB.  On a
more complex task like LMF, we see that \name is about 3
orders-of-magnitude faster than MADlib and \BDB's native tool.
This validates that \name is able to efficiently handle several
in-RDBMS analytics tasks, while offering a unified architecture.  
We also verified that all the tools compared achieved similar training
quality on a given task and dataset (recall that IGD converges to the optimal 
objective value on convex programs),
but do not present details here due to space constraints.

To understand why \name performs faster, we looked into the MADlib source code.
While the reasons vary across tasks, \name is faster generally because IGD has
lower time complexity than the algorithms in MADlib.
IGD, across all tasks, is linear in the number of examples (fixing the dimension)
and linear in the dimension of the model (fixing the number of examples).
But the algorithms in MADlib for LR, for instance, are super-linear in the dimension,
while that for LMF is super-linear in the number of examples.

To get a sense of the performance compared to other tools, a
comparison with the popular in-memory tool Weka shows that \name (over
PostgreSQL) is faster on all these tasks -- from 4X faster on dense LR
to over 4000X faster on dense SVM.  
We also validated that our runtimes on SVM are within a factor of 3X to the special-purpose SVM
in-memory tool, SVMPerf. This is not surprising as SVMPerf is highly
optimized for the SVM computation, but presents an avenue for future work.

\paragraph*{Next Generation Tasks}
Existing in-RDBMS analytics tools do not support emerging advanced analytics tasks like CRF. But \name is able to 
efficiently support even such next generation tasks within the same architecture. To validate this, we plot
the convergence over time for \name (over PostgreSQL) against in-memory tools. The results are shown in 
Figure \ref{tab:bench}(B). We see that \name is able to achieve similar convergence, and runtime as the hand-coded 
and optimized in-memory tools, even though \name is a more generic in-RDBMS tool.

\eat{
\begin{table}[h]
    \centering
    \begin{tabular}{|c||c||c|c|c|}
        \hline
        Task & Victor & MADlib & Weka & Others\\
        \hline
        \hline
        LR & $\surd$ & $\surd$ & XX & N/A\\
        \hline
        SVM & $\surd$ & X & XX & SVMPerf: XX\\
        \hline
        LMF & $\surd$ & X & X & N/A\\
        \hline
        \multirow{2}{*}{CRF} & \multirow{2}{*}{$\surd$} & \multirow{2}{*}{N/A} 
        & \multirow{2}{*}{N/A} & CRF++: XX\\
        & & & & Mallet: X\\
        \hline
    \end{tabular}
    \label{tab:scalability}
    \caption{Scalability of Tools: ($\surd$) scalable, (X) too long to finish, (XX) crashes due to memory error}
\end{table}
}

\begin{table}[h]
    \centering
    \begin{tabular}{|c||c||c|c|c|}
        \hline
        \multirow{2}{*}{Task} & \name & \ADB & \BDB& Others\\
        & PostgreSQL & (Native) & (Native) & (In-mem.)\\
        \hline
        \hline
        LR & $\surd$ & $\surd$ & $\surd$ & X\\
        \hline
        SVM & $\surd$ & $\surd$  & X & X\\
        \hline
        LMF & $\surd$ & N/A & X & X\\
        \hline
        CRF & $\surd$ & N/A & N/A & X\\
        \hline
    \end{tabular}
    \caption{Scalability : $\surd$ means the task completes, and X
      means that the approach either crashes or takes longer than 48
      hours. N/A means the task is not supported. The in-memory tools
      (Weka, SVMPerf, CRF++, Mallet) all either crash or take too
      long.}
    \label{tab:scalability}
\end{table}

\paragraph*{Scalability}
We now study the scalability of the various tools to much larger
datasets (Classify300M, Matrix5B and DBLP).  Since \name is not tied
to any RDBMS, we run it over PostgreSQL for this study. We compare
against the native analytics tools of both commercial engines, \ADB and \BDB,
as well as the task-specific
in-memory tools mentioned before.  The results are summarized in Table
\ref{tab:scalability}.  We see that almost all of the in-RDBMS tools
scale on the simple tasks LR and SVM (less than an hour per epoch for
\name), except \BDB on SVM, which did not terminate even after
48 hours.  Again, on the more complex tasks LMF and CRF, only \name
scales to the large datasets.  We also tried several custom in-memory
tools -- all crashed either due to insufficient memory (Weka, SVMPerf,
CRF++) or did not terminate even after 48 hours (Mallet).


\subsection{Impact of Data Ordering}
\label{sec:exp:ordering}

\begin{figure}[h]
\centering
\includegraphics[width=3.3in]{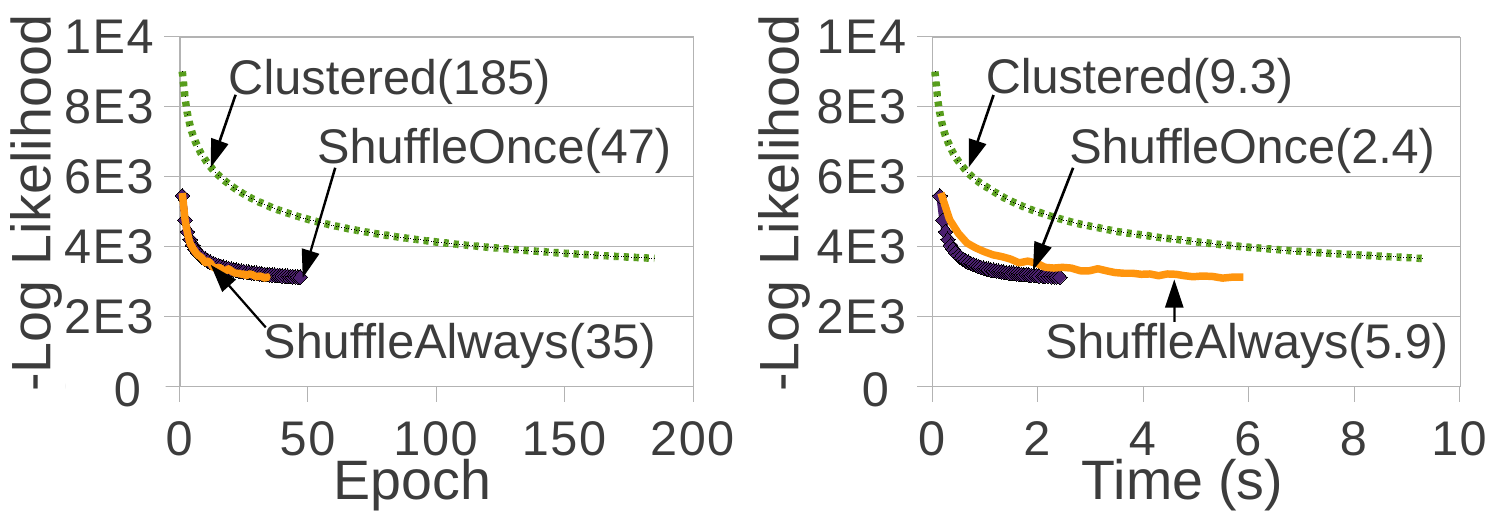}
\caption{Impact of Data Ordering on Sparse LR over DBLife: 
(A) Objective value over epochs, till convergence. 
The number of epochs for convergence are shown in parentheses.
(B) Objective value over time, till convergence. 
The time to converge (in sec) are shown in parentheses.}
\label{fig:shuffling}
\end{figure}

We now empirically verify how the order the data is stored affects the performance
of our IGD schemes. We first study the objective function value against epochs for 
data being shuffled before each epoch (ShuffleAlways). We repeat the study for data seen in clustered order 
(Clustered), without any shuffling. 
Finally, we shuffle the data only once, before the first epoch (ShuffleOnce). We present the results for 
the LR task on DBLife in Figure \ref{fig:shuffling}. We observed similar results on other datasets
and tasks, but skip them here due to space constraints.

Figure \ref{fig:shuffling}(A) shows that ShuffleAlways converges in the fewest epochs, as is expected for IGD. 
Clustered yields the poorest convergence rate, as explained in Section \ref{sec:ordering}. 
In fact, Clustered takes over 1000 epochs to reach the same objective value as ShuffleAlways. 
However, we see that ShuffleOnce achieves
very similar convergence rate to ShuffleAlways, and reaches the same objective value 
as ShuffleAlways in 12 extra epochs.
Figure \ref{fig:shuffling}(B) shows why the extra epochs are acceptable -- 
ShuffleAlways takes several times longer to finish than ShuffleOnce.
This is because the shuffling overhead is significantly high.
In fact, for simple tasks like LR, shuffling dominates the runtime -- e.g., for LR on DBLife, shuffling 
takes nearly 5X the time for gradient computation per epoch.
Even on more complex tasks, the overhead is significant, e.g., it is 
3X for LMF on MovieLens.
By avoiding this overhead, ShuffleOnce finishes much faster than ShuffleAlways, while still 
achieving the same quality.

\eat{
Figure \ref{fig:shuffling}(A) shows that shuffling always achieves the best convergence (in terms of epochs). This is in line
with what is expected for IGD. The clustered order yields the poorest convergence 
rate, as explained in Section \ref{sec:ordering}. However, we see that shuffling once achieves
very similar convergence to shuffling always, and reaches almost the same objective value within 
20 epochs.  Figure \ref{fig:shuffling}(B) shows that shuffling always takes much longer to finish than 
shuffle once.
For simple tasks like LR, shuffling dominates the runtime -- e.g., shuffling takes nearly 5X the time 
of gradient computation per epoch for LR on DBLife.
Though the overhead could be lower on more complex tasks, it is still significant -- e.g., it is 
3X for LMF on MovieLens.
By avoiding this overhead, shuffling once finishes much faster than shuffling always, while still 
achieving the same quality.
}

\subsection{Parallelizing IGD in an RDBMS}
\label{sec:exp:parallel}
\begin{figure}[h]
\centering
\includegraphics[width=3.3in]{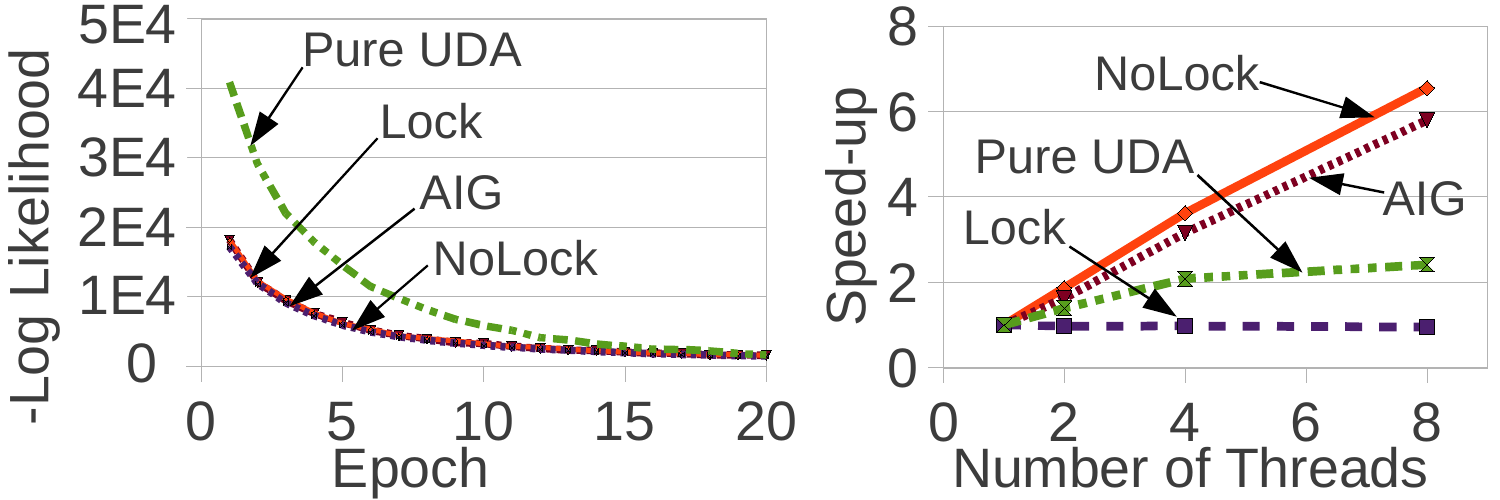}
\caption{Parallelizing IGD: (A) Plot of objective value over epochs for the pure UDA version and the 
shared-memory UDA variants (Lock, AIG, NoLock) for CRF over CoNLL on 8 threads (segments).
(B) Speed-up of the per-epoch gradient computation times against the number of threads. 
The per-epoch time of the single-threaded run is 20.6s.}
\label{fig:parallelism}
\end{figure}

We now verify that both the parallelism schemes (pure UDA and shared-memory UDA) are able to achieve 
near-linear speed-ups but the pure UDA has a worse convergence rate than the shared-memory UDA.
We first study the objective value over epochs for both the implementations. We use 
the three concurrency schemes for the shared-memory UDA -- lock the model (Lock), AIG, and 
no locking (NoLock). We present the results for CRF on CoNLL in Figure \ref{fig:parallelism}(A) 
(similar results on other tasks skipped here for brevity).

Figure \ref{fig:parallelism}(A) shows that the pure UDA implementation 
has poorer convergence rate compared to the shared-memory UDA with Lock, since the 
model averaging in the former yields poorer quality \cite{DBLP:conf/icdm/ZhuCWZC09}. The figure also shows that AIG and NoLock 
have similar convergence rate to the Lock approach. This is in line with recent results from the machine learning literature 
\cite{hogwild}. By adopting the NoLock shared-memory UDA parallelism into \name, we achieve significant speed-ups in a 
generic way across all the analytics tasks we handle. 
Figure \ref{fig:parallelism}(B) shows the speed-ups (over a single-threaded run) achieved 
by the four parallelism schemes in \BDB. As expected, the Lock approach has no speed-up, while 
the speed-up of the pure UDA approach is sub-optimal due to model passing overheads. NoLock and AIG achieve linear speed-ups, with
NoLock having the highest speed-ups.


\subsection{Multiplexed Reservoir Sampling}
\eat{
TODO: graph of loss vs epoch for clust, subsampl and resvr;\\
TODO: table of time to 2x opt with B , subsmpl , resvr\\
Both on dblife dataset for sparse lr.\\
}
We verify that our Multiplexed Reservoir Sampling (MRS) scheme has faster
convergence rate compared to both Subsampling and operating over clustered data (Clustered).

\begin{figure}[h]
\centering
\begin{tabular}{cc}
\parbox{1.35in}{
\includegraphics[width=1.5in]{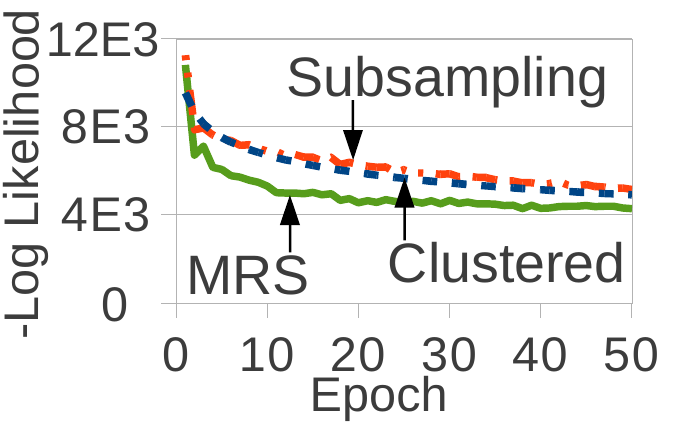}
}
&
\begin{tabular}{|c||c|c|}
\hline
\multirow{2}{*}{B} & Sub-& \multirow{2}{*}{MRS}\\
& Sampling &\\
\hline
\hline
800 & 2.50 (48) & 0.60 (10)\\
\hline
1600 & 1.37 (26)& 0.36 (6)\\
\hline
3200 & 0.69 (13) & 0.12 (2)\\
\hline
\end{tabular}
\end{tabular}
\caption{Multiplexed Reservoir Sampling: (A) Objective value against epochs for LR on DBLife.
The buffer size for Subsampling and MRS is 1600 tuples (10\% of the dataset). 
(B) Runtime (in sec) to reach 2X the optimal objective value for different buffer sizes, B. The 
numbers in parentheses indicate the respective number of epochs. The same values for Clustered 
are 1.03s (19).}
\label{fig:reservoir}
\end{figure}

Figure \ref{fig:reservoir}(A) plots the objective value against epochs for the three schemes.
For Subsampling and MRS, we choose a buffer size that is about 10\% the 
dataset size (for LR on DBLife). We see from the figure that MRS
has faster convergence rate than both Subsampling and Clustered, and 
reaches an objective value that is 20\% lower than both. 
Figure \ref{fig:reservoir}(B) shows the sensitivity to the buffer size for the Subsampling
and MRS schemes. We see that the runtime to reach 2X of the optimal objective value
is lower for MRS. This is as expected since MRS has faster convergence rate than Subsampling.
Finally, we verify that \name with the MRS scheme provides better 
performance than existing in-RDBMS tools on large datasets (that do not fit in available RAM).
For a simple task like LR on the Classify300M dataset over PostgreSQL, 
with a buffer that is just 1\% of the dataset size, \name with the MRS scheme
achieves the same objective value as MADlib in 45 minutes, while MADlib takes over 3 hours.
On a more complex task like LMF on the Matrix5B dataset, 
\name with MRS scheme finishes in a few hours, while MADlib
did not terminate even after one week.

\section{Conclusions and Future Work}
We present \name, a novel architecture that takes a step towards
unifying in-RDBMS analytics. Using insights from the mathematical
programming literature, \name provides a single systems-level
abstraction to implement a large class of existing and next-generation
analytics techniques.  In providing a unified architecture, we argue
that \name may reduce the development overhead for introducing and
maintaining sophisticated analytics code in an RDBMS.  \name also
achieves high performance on these techniques by effectively utilizing
standard features available inside every RDBMS.  We implemented \name
over two commercial RDBMSes and PostgreSQL, and verified that \name
achieves competitive, and often superior, performance than the
state-of-the-art analytics tools natively offered by these RDBMSes.

While \name can handle many analytics techniques in the current
framework, it is interesting future work to integrate more
sophisticated models, e.g., simulation models, into our architecture.
Another direction is to handle large-scale combinatorial optimization
problems inside the RDBMS, including tasks like linear programming and
fundamental \textsf{NP}-hard problems like MAX-CUT.

One area to improve \name is to match the performance of some
specialized tools for tasks like support vector machines by 
using more optimizations, e.g. model or feature compression.  There
are also possibilities to improve performance by modifying the DBMS
engine, e.g., exploiting better mechanisms for model passing and
storage, concurrency control, etc. Another direction is to examine
more fully how to utilize features that are available in parallel
RDBMSes.

\section{Acknowledgments}
This research has been supported by the ONR grant N00014-12-1-0041, 
the NSF CAREER award IIS-1054009, and gifts from EMC Greenplum and Oracle 
to Christopher R\'{e}, and by the ONR grant N00014-11-1-0723 to Benjamin Recht.
We also thank Joseph Hellerstein, and the analytics teams from EMC Greenplum 
and Oracle for invaluable discussions.

\submissionversion{\balance}

\bibliographystyle{plain} 
\bibliography{chris.local.wopages} 

\appendix
\section{Proximal Point Methods}
\label{a:grads}

\eat{\begin{figure}
\centering
{\small
\begin{tabular}[t]{|c|c|}
\hline
$P(x)$ & $\Pi_{\alpha P}(z)$\\
\hline
$\lambda\|x\|_2$ & $(1+\alpha\lambda)^{-1}z$ \\
$\mu \|x\|_1$ & $\mathrm{sgn}(z)\circ\max(|z|-\alpha\mu,0)$ \\
$-\sum_{i=1}^n \ln(x_i)$ & $\tfrac{1}{2}\left( z+\sqrt{z^2+4\alpha}\right)$ \\
\hline
\hline
Constraint & $\Pi_{\alpha P}(z)$\\
\hline
$\|x\|_2\leq B$ & $z/\max(B,\|z\|_2)$\\
$\max_j \|x_j\|_2 \leq B$ &  $z_j/\max(B,\|z_j\|_2)$ for all $j$\\
$x\geq 0$ & $\max(z,0)$\\
\hline
\hline
\hline
Objective & Gradient\\
\hline
$\tfrac{1}{2}(a^Tx-b)^2$ & $(a^Tx-b)a$ \\
$(1-yx)_+$ & $-\mathbf{1}_{\{yx<1\}} yx$\\
$\log(1+\exp(-yx))$ & $-(1+\exp(yx))^{-1}yx$\\
\hline
\end{tabular}}
\caption{Gradients and Proximal Operators. \scriptsize
  Here, $\mathbf{1}_S$ denotes the function equal to $1$ if $x\in S$
  and zero otherwise.}
\label{fig:prox}
\end{figure}
}

To handle regularization and constraints, we need an additional
concept called {\em proximal point methods}. These do not
change the data access patterns, but do enable us to handle
constraints. We state the complete step rule including a projection
that allows us to handle constraints:

\begin{equation}
 w^{(k+1)} = \Pi_{\alpha P}\left(w^{(k)} - \alpha_k \nabla f_{\eta(k)}(w^{(k)})\right)
\label{eq:rule:extended}
\end{equation}

Where the function $\Pi_{\alpha P}$ is called a proximal point
operator and is defined by the expression:
\[
	\Pi_{\alpha P}(x) = \arg\min_w \quad \tfrac{1}{2} \|x-w\|_2^2 + \alpha P(w)
\]
In the case where $P$ is the indicator function of a set $C$,
$\Pi_{\alpha P}$ is simply the Euclidean projection onto
$C$~\cite{Rockafellar76}. Thus, these constraints can be used to
ensure that the model stays in some convex set of constraints. An
example proximal-point operator ensures that the model has unit
Euclidean norm by projecting the model on to the the unit ball.
$P(w)$ might also be a regularization penalty such as total-variation
or negative entropy. These are very commonly used in statistics to
improve the generalization of the model or to take
advantage of properties that are known about the model to reduce the
number of needed measurements. \eat{Examples of gradients and proximal
point methods are given in Figure~\ref{fig:prox}.}

\section{Background: Step-size and Stopping Condition}
\label{a:step}
\eat{
The divergent series rule, anstreicher (two-well known facts about the
subgradient method). Reproduce the proof here for completeness.

In practice people use constant step-size and number of epochs.

Explain why gradient size is a good termination condition for strongly
convex functions.
}
The step-size and stopping condition are the two important rules for
gradient methods. In real-world systems, constant step-sizes and fixed number of
epochs are usually chosen by an optimization expert and set in the software
for simplicity. 
In some cases, number of epochs or tolerance rate are exposed to end users as parameters. 

Theoretically, to prove that gradient methods converge to the optimal value, 
it requires step-sizes to satisfy some properties. For example, the proof for
divergent series rule:
\[
  \alpha_k \rightarrow 0, \sum_{k=1}^{\infty}\alpha_k = \infty, 
\]
and geometric rule:
\[
  \alpha_k = \alpha_0 \rho^{k}, 0 < \rho < 1, \alpha_0 > 0,
\]
are given in Anstreicher \cite{DBLP:journals/mp/AnstreicherW09}. For
strongly convex objective functions, the distance between a point $x$
and the optimal value $x*$ can be bound by $||\nabla f(x)||$ which
provides a more rigorous stopping condition. In our architecture, we
can support all of the above rules.

\section{Calculations for CA-TX Example}
\label{a:txca}
\newcommand{\E}{\mathbb{E}}
Suppose we start from $w_0$ and we run for $m$ iterations. Then the
behavior of any IGD algorithm can be modeled as a function $\sigma : m
\to n$, i.e., $\sigma(i) = j$ says that at step $i$ we picked example
$j$. Let us assume a constant step-size $\alpha \geq 0$. Then, the IGD
dynamic system for the example is:
\[ w_{k+1} =  w_k - \alpha (w_k - y_\sigma(k)) \]
We can unfold this in a closed form to:
\[ w_{k+1} = (1-\alpha)^{k+1} w_0 + \alpha \sum_{j=0}^{k} (1-\alpha)^{k-j} y_{\sigma(j)} \]
From this, we can see that the graphs shown in the example are not
random chance. Specifically, we can view $\sigma(i)$ for $i=1,\dots,m$
as a random variable. For example, suppose that $\sigma$ models
selecting without replacement then observe that, $\Pr[y_{\sigma(i)} =
  1] = \Pr[y_{\sigma(-i)} = -1] = 1/2$. Said another way, this
sampling scheme is unbiased. We denote by $\E_{wo}$ the expectation
with respect to a without replacement sample (assume $m \leq 2n$ for
simplicity). From here, one can see that in expectation
$w_{k+1}$ goes to $0$ as expected. One can see that
the convergence holds for any unbiased scheme. 

For the deterministic order in the CA-TX example, we have $\sigma(i) =
i$. And recall, that we intuitively converge to $-1$ (since $y_i = -1$
for $i \geq n$):
\begin{align*}
w_{2n} & = (1 - \alpha)^{2n} w_0 - \alpha (1 - (1-\alpha)^{n}) \frac{1 - (1-\alpha)^{n+1}}{1- \alpha} \\
  & = (1 - \alpha)^{2n} w_0 - (1 - (1-\alpha)^{n})^2 - \alpha (1-\alpha)^{n}
\end{align*}

Indeed if $\alpha$ is large so that $(1-\alpha)^{n} \approx 0$, then
we converge to roughly $-1$. If however, $(1-\alpha)^{n}$ is very close
to $1$ then the initial condition matters quite a bit. Of course, as
$\alpha$ decays it passes through a sweet spot where it eventually
converges to $1$ after a few epochs. It is not hard to see the
stronger statement that the deterministic example is a worst case
ordering for convergence (the other is $\sigma(i) = 2n - i$).

\end{document}